\documentclass[floatfix]{iopart}
\usepackage[sort&compress,square,comma,numbers]{natbib}
\usepackage{graphicx}
\usepackage{color}
\usepackage{amssymb}

\newcommand{\eqref}[1]{(\ref{#1})}

\begin{document}
\title{Low-energy properties of the Kondo lattice model}
\author{O Bodensiek$^1$, R \v{Z}itko$^2$, R Peters$^3$ and T Pruschke$^1$}
\address{Institute for Theoretical Physics, University of G\"ottingen, 
Friedrich-Hund-Platz 1,
37077 G\"ottingen, Germany}
\address{J.\ Stefan Institute, Jamova 39, SI-1000 Ljubljana, Slovenia}
\address{University of Kyoto, Department of Physics, Graduate School
  of Science,  Kitashirakawa-Oiwakecho, 606-8502 Kyoto, Japan}
\ead{pruschke@theorie.physik.uni-goettingen.de}

\begin{abstract}
We study the zero-temperature properties of the Kondo lattice model 
within the dynamical mean-field theory.
As impurity solver we use the numerical renormalization group. 
We present results
for the paramagnetic case showing the anticipated heavy Fermion physics, including
direct evidence for the appearance of a large Fermi surface for antiferromagnetic
exchange interaction. 
Allowing for the formation of a N\'eel state, we observe at finite doping an
antiferromagnetic metal below a critical exchange interaction, which shows
a crossover from a local-moment antiferromagnet with a small Fermi surface for weak
exchange coupling to a heavy-fermion antiferromagnet with a large Fermi surface
for increasing exchange. 

Including lattice degrees of freedom via an additional Holstein term 
we observe a significant suppression of the Kondo effect, leading to strongly reduced low-energy scale. For too large electron-phonon coupling we find a complete collaps of the heavy Fermi liquid and the formation of polarons.

%By introducing a Nambu notation, we find that increasing electron-phonon
%coupling favors superconductivity, which however is not BCS like but shows
%additional structures in the density of states and the gap function.
%Due to the latter there is a distinction between the mean-field gap and the gap observed in the tunneling density of states.
\end{abstract}
\submitto{\JPC}
%\maketitle              

\section{Introduction}
Heavy Fermion systems based on 4f or 5f intermetallics
are paradigms for electronic correlations in solid state physics. 
The low-temperature physics of these
compounds is strongly influenced by the local moment on the f shell, subject 
to an antiferromagnetic exchange to the conduction electrons. The 
resulting physical properties are in many cases again Fermi liquid like,
however with extremely enhanced Landau parameters, in particular an effective mass
up to three orders of magnitude larger than the one found in conventional metals\cite{stewart:1984,GreweSteglich:1991,stewart:2001}.
This large effective mass is the reason why these systems are referred to as
\emph{Heavy-Fermion materials} (HF). 
Moreover, in addition to these extreme Fermi liquid properties, the HF also
show various phase transitions and the thermodynamics show that these
transitions actually occur within the heavy Fermi liquid \cite{stewart:1984}.
Finally, the appearance of superconductivity in a system with initially
well-defined magnetic moments is a rather unconventional feature, and close
investigation revealed early on that the nature of the ordered state may be
rather unconventional \cite{GreweSteglich:1991}.
This observation has been substantiated by the development over the past 15 years
which showed that a larger number of these HF systems exhibit rather
peculiar quantum-phase transitions, partially identified as the driving force behind
the superconducting transitions \cite{stewart:2001,loehneysen:2007}.

% The Kondo-lattice model (KLM) applies to systems where delocalized 
% quasi-particle states, describing the conduction electrons,
% interact with strongly localized electrons or spins via a weak hybridization or
% an exchange interaction. Such a situation typically appears in compounds 
% involving elements from the rare-earths. 

Other materials which show a coupling between itinerant quasi-particles and
localized spins are 
certain transition metal oxides \cite{imada:98}, magnetic semiconductors or 
semi-metals in the series of the rare earth monopnictides and monochalcogenides
\cite{ovchinnikov:91,sharma:06},
and diluted magnetic semiconductors such as Ga$_{1-x}$Mn$_x$As
\cite{ohno:98,dms_review}. Here, the coupling between local spin and conduction electrons 
usually mediated through Hund's exchange and thus typically is ferromagnetic.

A theoretical description of HF compounds is conventionally based on the
Kondo-lattice model (KLM)
\begin{equation}\label{eq:KLM}
H_{\rm KLM}=\sum_{\vec{k},\sigma}\epsilon_{\vec{k}}\hat{c}_{\vec{k},\sigma}^\dagger \hat{c}_{\vec{k},\sigma}-J\sum_i\vec{s}_i\cdot\vec{S}_i\;\;.
\end{equation}
The operators $\hat{c}_{\vec{k},\sigma}^{(\dagger)}$ denote annihilation (creation) 
operators of itinerant quasi-particles with dispersion $\epsilon_{\vec{k}}$,
$\vec{s}_i$ is the operator for the conduction states' spin density at
lattice site $\vec{R}_i$ and $\vec{S}_i$ describes a spin of magnitude $S$
localized at site $\vec{R}_i$. The interaction between the spin of the conduction
states and the localized spin is modeled as conventional isotropic exchange 
interaction $J$.

The dilute version of the KLM \eqref{eq:KLM}, the so-called
single-impurity Kondo model (SIKM) where there exists only one additional
spin at site $\vec{R}_i=0$, is well understood and
shows for antiferromagnetic coupling $J<0$ the Kondo effect \cite{hewson:book},
which precisely leads to the phenomena observed in the Fermi liquid phase of 
HF systems, viz a 
strongly enhanced mass. There are nowadays
several computational tools to treat the SIKM, for instance continuous-time Monte-Carlo \cite{Otsuki:2007}
or Wilson's numerical renormalization group \cite{bullareview}.

In theoretical treatments one usually ignores the lattice degrees
of freedom. On the other hand, all the above mentioned materials have a rather
strong electron-phonon coupling \cite{allen:1982,imada:98} and one can expect
that the charge physics driven by phonons somehow competes with the spin
physics due to the exchange interaction with the localized spin. Moreover,
without exchange coupling, phonons will lead to conventional $s$-wave
superconductivity. Thus the investigation of the interplay between 
a coupling to a spin and the lattice degrees of freedom is highly interesting.

This paper is intended to give a summary of the physical properties of the
KLM as seen by the dymanical mean-field theory. This necessesarily excludes
nonlocal phenomena like unconventional superconductivity, but allows for
the study of antiferromagnetism and whether it is always accompanied by
a breakdown of the large Fermi surface. Inclusion of phonons eventually
leads to superconductivity \cite{Bodensiek:2009}, which however is of the 
standard local $s$-wave type. A detailed account of an investigation of the
interplay of phonon-mediated superconductivity and HF physics
will be presented elsewhere.

The paper is organized as follows. In the next section we discuss the model
and the approximation used to solve it. The case without phonons, i.e.\ the
conventional heavy-fermion physics both in the paramagnetic and the antiferromagnetically
ordered state
is the subject of section \ref{sec:parm_metal}. 
The effect of phonons on the low-energy properties 
%Introducing a Nambu formalism
%allows to study superconductivity, which 
will be discussed in section \ref{sec:phonons}.
%section \ref{sec:sc}.
A short summary and outlook will close the paper.

\section{Model and Method}
The KLM Hmiltonian \eqref{eq:KLM} will be again the basic model.
Except for one dimension, no analytical solution exists, and even conventional
numerical tools such as Quantum Monte-Carlo (QMC) become rather cumbersome due
to a severe sign problem away from particle-hole symmetry.
Thus, a reliable approximate method is needed. If one is not interested in 
the properties too close to a phase transition or in the rather complicated, non-local
ordering phenomena, a suitable tool is the dynamical mean-field theory (DMFT)
\cite{georges:96}.
Here, the lattice is mapped onto an effective single-impurity problem, which
can be then solved using standard techniques. Here, we use the numerical renormalization
group (NRG) approach \cite{wilson:75,bullareview}.
One of its apparent advantages is the possibility to access small energy scales
without problem and cover the whole range from $T=0$ to finite temperatures of
the order of the bare energy scales. Furthermore, it also allows to
include phonons to a certain extent, namely an Einstein mode coupled through a
Holstein term to the charge degrees of freedom. As is suggestive from effects
like Kondo volume collapse \cite{allen:1982,imada:98}, such a term could be 
rather important. We will thus work with a Hamiltonian (for a detailed
introduction and further references see \cite{assaad:2009})
\begin{equation}
  \label{eq:KLM_Holstein}
  H=H_{\rm KLM}
+
\omega_0\sum\limits_{i}b^\dagger_i b^{\phantom{\dagger}}_i
+g\sum\limits_{i\sigma}\left(
c_{i\sigma}^\dagger c_{i\sigma}^{\phantom{\dagger}}-1\right)\left(b^\dagger_i
+b^{\phantom{\dagger}}_i\right)
\end{equation}
where $\omega_0$ is the frequency of an appropriate optical mode and $g$ a 
measure of the electron-phonon coupling. How such an additional local 
coupling can be treated within NRG, is described in detail in \cite{bullareview}.
To obtain reasonably accurate spectra also for higher energies, we use
the broadening strategy introduced by \cite{Freyn:2009}.

As is well-known, one major effect of the Holstein model is to introduce an
effective attractive interaction to the electronic subsystem: If the phonon
frequency $\omega_0$ and the 
electron-phonon coupling $g$ become large, keeping $g^2/\omega_0$ constant,
the phonons can be integrated out, yielding an attractive local Coulomb
interaction $U_{\rm eff}=-2g^2/\omega_0$. Without explicit exchange interaction
$J$, one will then obtain a Hubbard model with attractive $U$, which shows
charge-density and superconducting ordering phenomena
\cite{Freericks:1993,Bauer:2009}.
From the point of view of Kondo physics, the negative $U$ will lead to a Kondo-like
behavior in the charge sector, strongly competing with the spin Kondo effect
introduced by $J$. We thus can expect interesting physics to occur
when both couplings are present.

\section{KLM without phonons\label{sec:parm_metal}}
Let us begin with 
a comprehensive summary of the basics of heavy-fermion physics. 
We will use a $2D$ square lattice with nearest-neighbor hopping
for the conduction states. Note that
the DMFT is rather insensitive to the dimensionality, and we chose the $2D$ lattice
to facilitate visualization of the results.
Calculations were done with an NRG discretization parameter $\Lambda=2$, 
between 1000 and 5000 states were kept per NRG step and, where applicable, 
50 bosons kept initially. These values were systematically
changed for selected calculations
to ensure that the results are independent of these numerical parameters.
\subsection{Paramagnetic metal}
Let us start with a comparison 
of the properties at finite $J$, but with $g=0$. We can distinguish two cases,
namely an antiferromagnetic exchange $J<0$ and a ferromagnetic $J>0$.
The resulting density of states (DOS) for $T=0$
\begin{figure}[htb]
  \centering
  \includegraphics[width=0.8\textwidth]{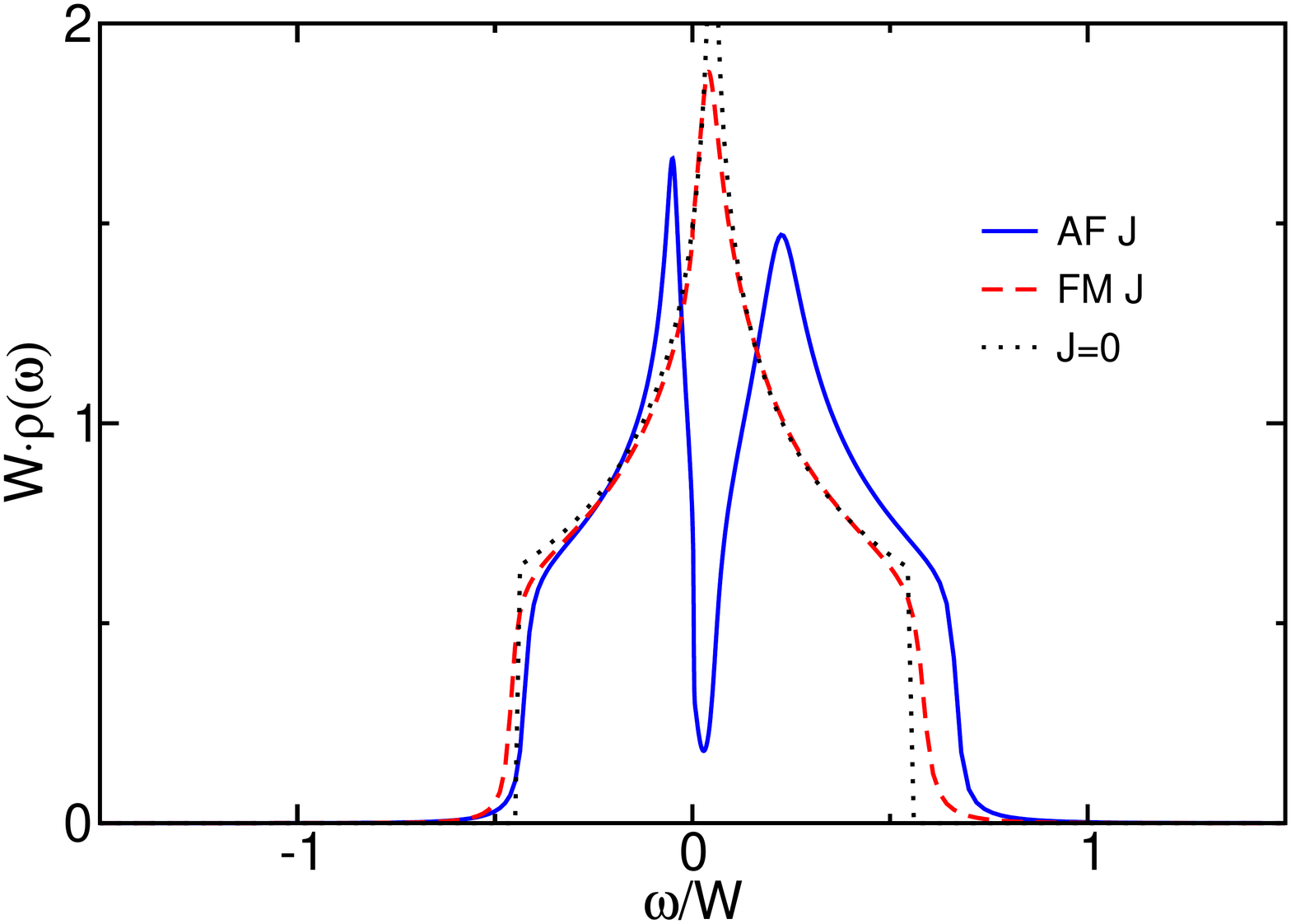}
  \caption{DOS at $T=0$ for the KLM \eqref{eq:KLM} with $|J|=0.25W$, where $W$ denotes
the bandwidth of the conduction electrons. The filling of the band is $n_c=0.8$. The dotted curve displays the DOS at $J=0$ as reference.}
  \label{fig:DOS_no_phonons}
\end{figure}
and $|J|/W=0.25$ at a filling $n_c=0.8$ of the
conduction band is shown in Fig.\ \ref{fig:DOS_no_phonons}. The quantity $W$
denotes the bandwidth of the conduction band and will serve as energy scale
hereafter. There are notable differences between the two cases
$J<0$ and $J>0$. The DOS for $J>0$ looks very much like the
DOS of the bare conduction band (dotted curve in Fig.\ \ref{fig:DOS_no_phonons}), 
although it is somewhat broadened.
\begin{figure}[htb]
  \begin{minipage}{0.45\textwidth}
    \begin{center}
      \includegraphics[width=\textwidth]{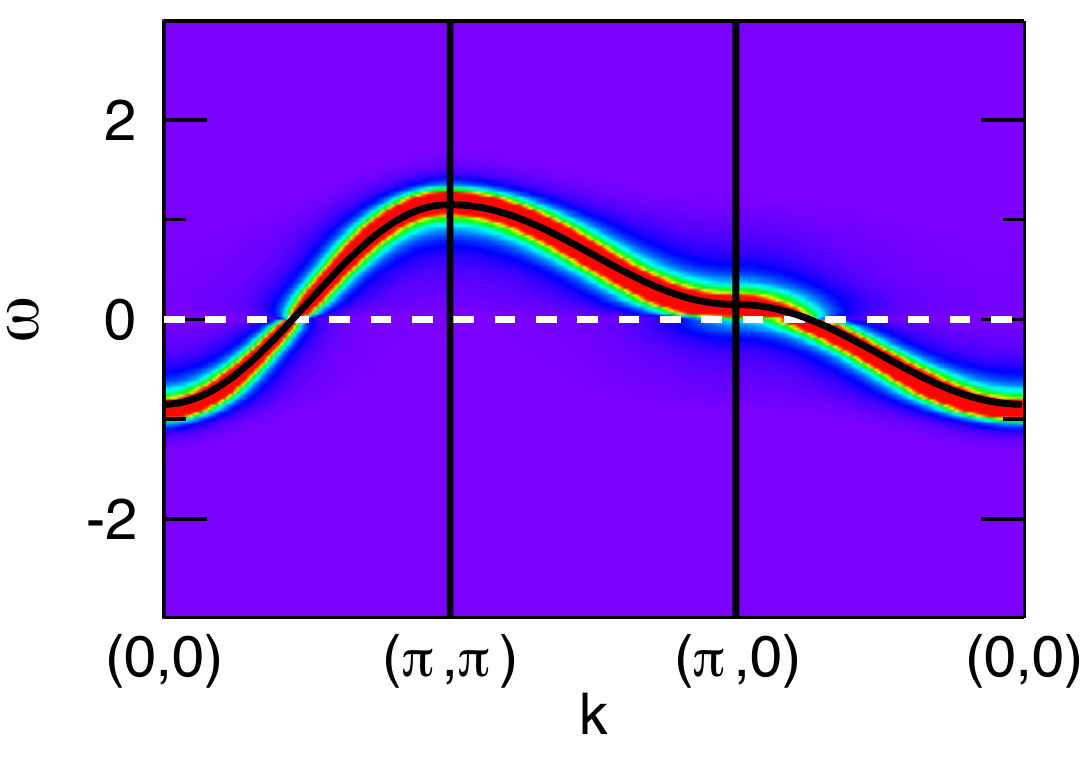}
    \end{center}
  \end{minipage}
  \hfill
  \begin{minipage}{0.45\textwidth}
    \begin{center}
      \hspace*{-5mm}\includegraphics[width=1.25\textwidth]{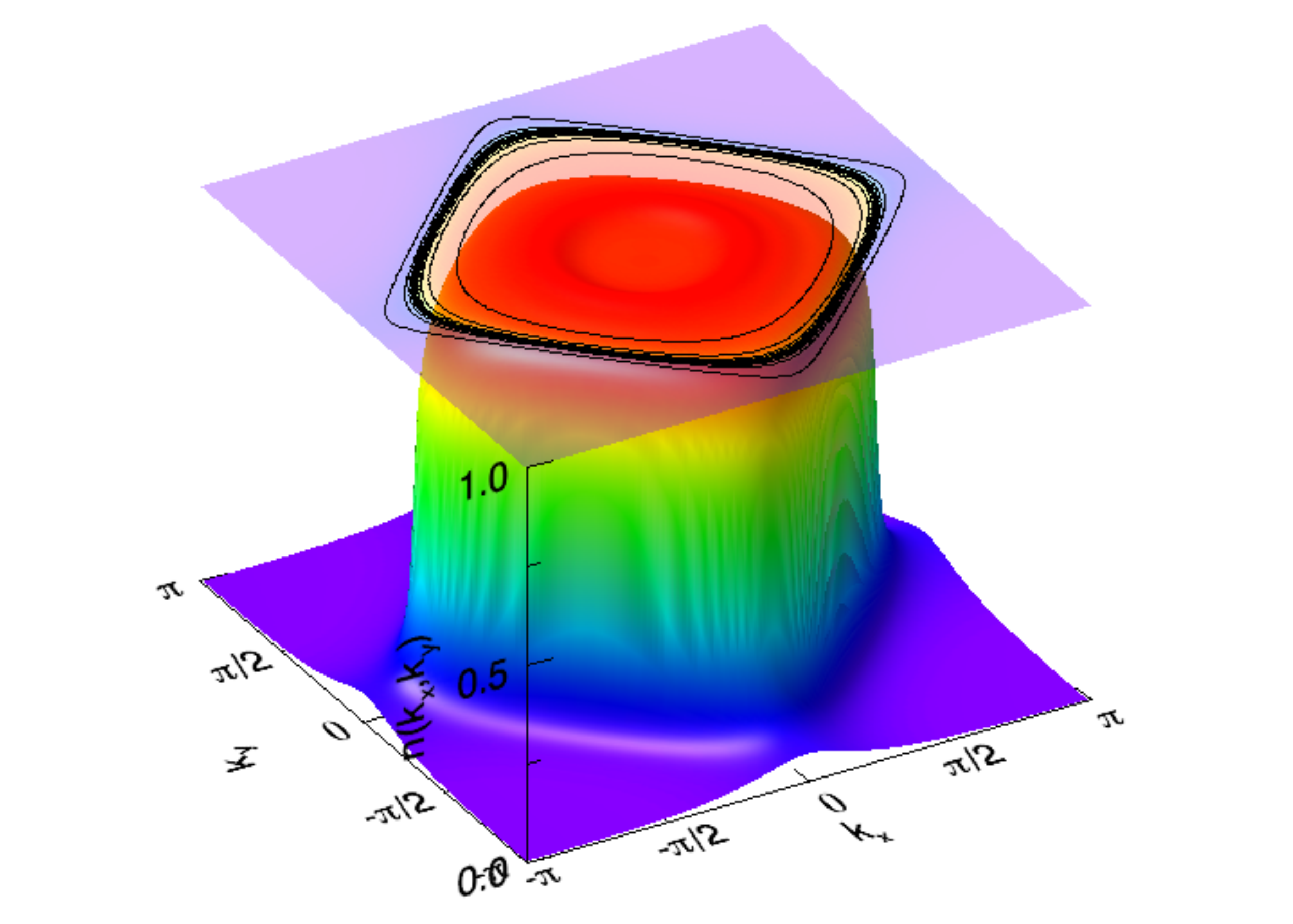}
    \end{center}
  \end{minipage}\\[-2cm]
 \centerline{$J>0$}
 \mbox{}\\[2cm]
  \begin{minipage}{0.45\textwidth}
    \begin{center}
      \includegraphics[width=\textwidth]{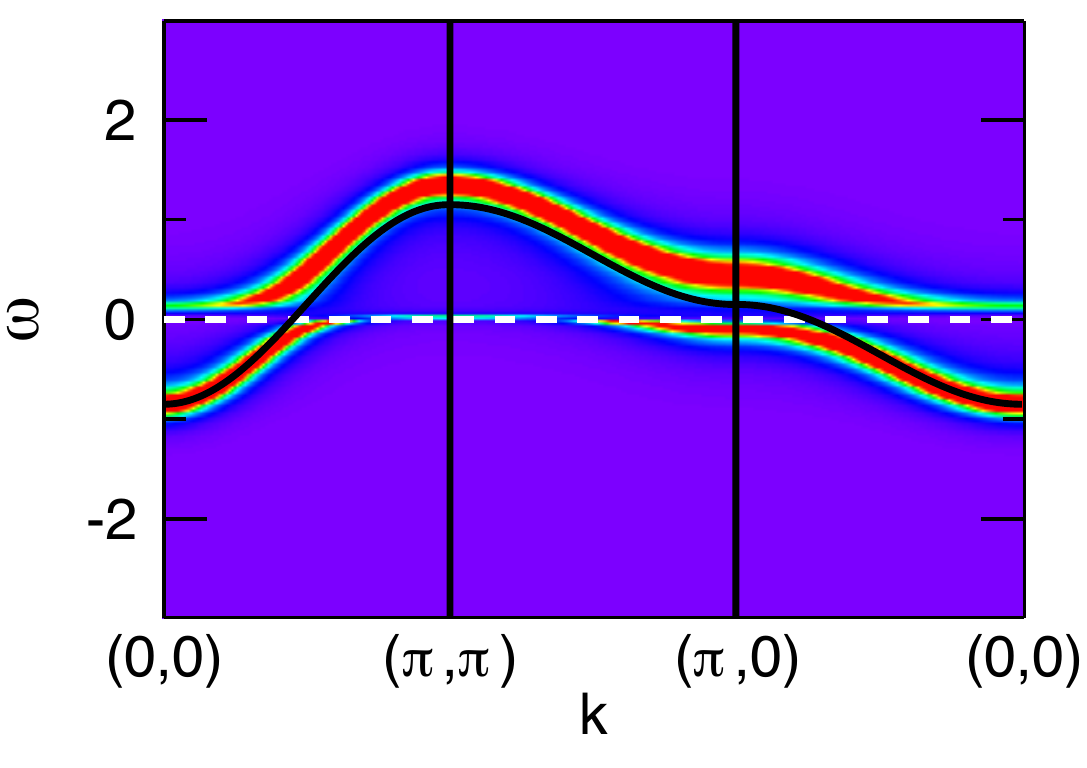}
    \end{center}
  \end{minipage}  
  \hfill
  \begin{minipage}{0.45\textwidth}
    \begin{center}
      \hspace*{-5mm}\includegraphics[width=1.25\textwidth]{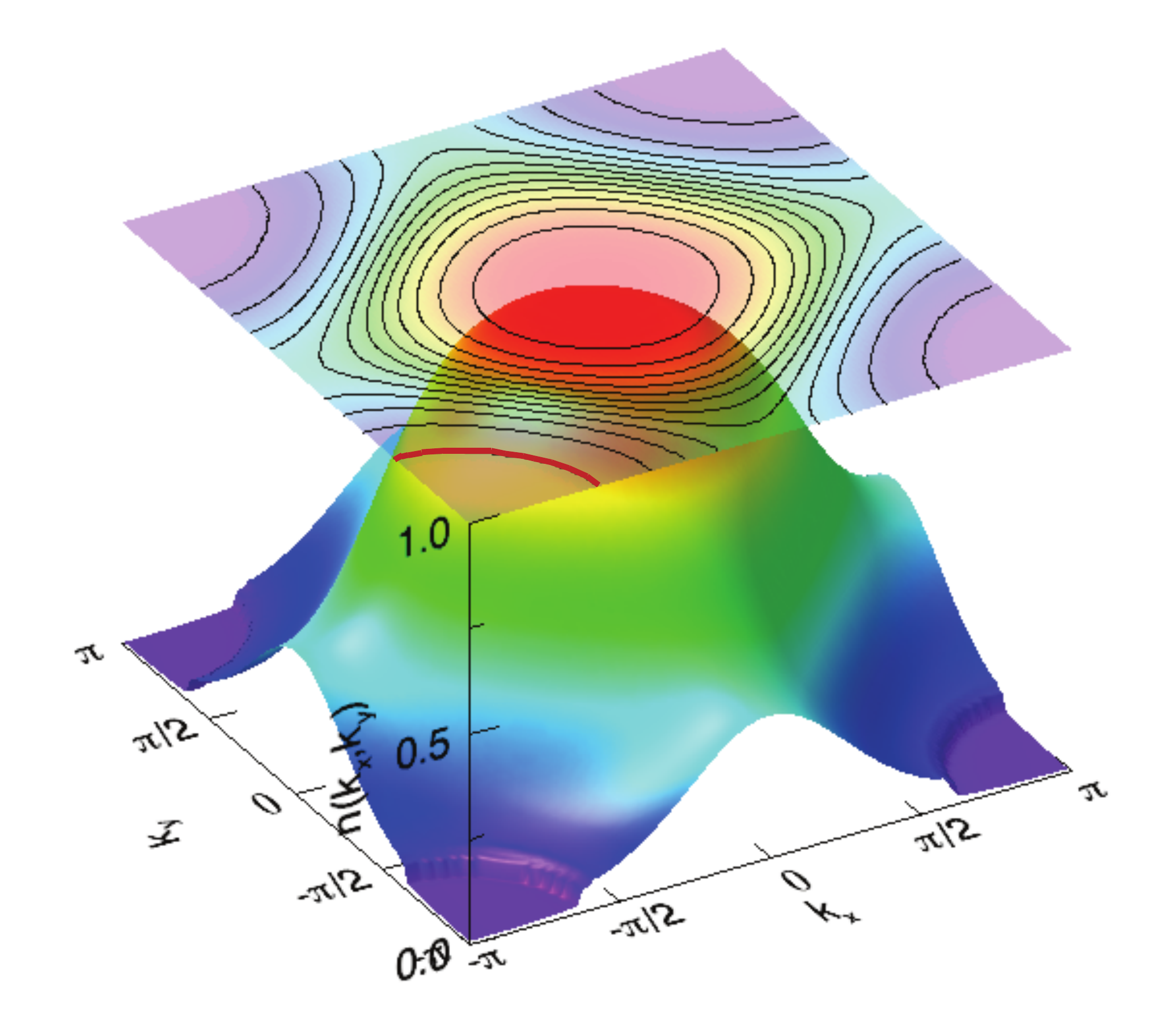}
    \end{center}
  \end{minipage}\\[-2cm]
  \centerline{$J<0$}
  \mbox{}\\[1cm]
  \caption{Spectral functions (left) and momentum distribution (right)
    for $|J|=0.25W$, other parameters as in Fig.\ \ref{fig:DOS_no_phonons}. The thick line in the lower left corner of the contour plot for $J<0$ denotes the position of the jump in the momentum distribution.}
  \label{fig:Aofkw_no_phonons}
\end{figure}
For $J<0$, however, the DOS is strongly modified \cite{otsuki:2009}, showing a pseudo gap close
to the Fermi energy $\omega=0$. This latter feature is a fingerprint of heavy
Fermion physics, resulting from a picture of hybridized bands
\cite{GreweSteglich:1991}.
This interpretation becomes even more apparent when one looks at the 
spectral function along the standard $\vec{k}$ directions in the first
Brillouin zone of the $2D$ square lattice in Fig.\ \ref{fig:Aofkw_no_phonons},
left part. Dark color means low intensity, bright color high.
Included as full black line is the bare bandstructure of the $2D$ nearest-neighbor
tight-binding band.
For $J>0$ we basically see the band structure of the $2D$ nearest
neighbor tight-binding band. There is a moderate broadening, which actually
is to be expected, because for $J>0$ the local spin effectively acts as a
potential scatterer for the band states \cite{hewson:book}. 
For such a situation the DMFT is equivalent to a CPA 
calculation, which yields a constant broadening.
% \begin{figure}[htb]
%   \begin{minipage}[t]{0.45\textwidth}
%     \begin{center}
%       $J>0$\\
%       \hspace*{-5mm}\includegraphics[width=1.25\textwidth]{Nofk_no_phonons_FMJ}
%     \end{center}
%   \end{minipage}
%   \hfill
%   %\hspace*{-5mm}
%   \begin{minipage}[t]{0.45\textwidth}
%     \begin{center}
%       $J<0$\\
%       \hspace*{-5mm}\includegraphics[width=1.25\textwidth]{Nofk_no_phonons_AFJ}
%     \end{center}
%   \end{minipage}
%   \caption{Momentum distribution $n(\vec{k})$ for $|J|=0.25W$, other parameters as in Fig.\ \ref{fig:DOS_no_phonons}.}
%   \label{fig:Nofk_no_phonons}
% \end{figure}
For $J<0$ (right panel), on the other hand,  there is a flattening of the
band structure close to the Fermi energy and a structure similar to a hybridization
gap opens. The flat portion of the lower band corresponds to a large effective 
mass of the quasi-particles. Note that we have a rather sharp structure at the
Fermi energy, i.e.\ one can indeed talk about quasi-particles here.

Another remarkable difference between the cases $J>0$ and $J<0$ is
observed when one looks at the momentum distribution function $n(\vec{k})$
displayed in the right part of Fig.\ \ref{fig:Aofkw_no_phonons}.
As already noted for the spectral functions,
the result for $J>0$ resembles the Fermi function, slightly smeared out by
incoherent scattering from the spins. In any case, the Fermi surface is located
at the $\vec{k}_F$ of a non-interacting $2D$ tight-binding band with a filling
$n_c=0.8$. On the other hand, the momentum distribution for $J<0$ does not show
any distinct features
at this particular value of $\vec{k}$. Instead, one notes a small jump in $n(\vec{k})$
at a vector outside the square marking the Fermi surface of a half-filled system.
A closer inspection shows that this $\vec{k}$ vector corresponds to a Fermi
surface of a system with $n_c=1.8$, i.e.\ the system shows a ``large'' Fermi
surface with the spin degrees of freedom contributing to the quasi-particles
now. Note that the height of the jump in $n(\vec{k})$ is directly related
to the inverse effective mass of the quasi-particles.

The results for $J<0$ in Fig.\ \ref{fig:Aofkw_no_phonons} represent
the essence of HF physics, namely the generation of heavy quasi-particles,
represented by flat bands with a structure known from hybridized bands and
a large Fermi surface.
\subsection{Antiferromagnetic ordering}
As is well known, the Kondo lattice model shows a variety of magnetically ordered
states \cite{lacroix:1979,fazekas:1991,kienert:2006,henning:2009}. Within DMFT, an overview was presented in \cite{Peters:2007,chattopadhyay:2001}. The
% \begin{figure}[htb]
%   \centering
%   \includegraphics[width=0.95\textwidth]{maghalf}
%   \caption{Staggered magnetization at half filling and $T=0$ as function of
% exchange coupling $J$.}
%   \label{fig:MagHalf}
% \end{figure}
major findings at $T=0$ are:
% collected in Fig.\ \ref{fig:MagHalf}: 
At half filling (Kondo insulator regime) one finds a
critical $J_c<0$, with no magnetic phases present for $J<J_c$ and 
antiferromagnetism for $J>J_c$. 
The staggered magnetization of local spins and band electrons are 
opposite for $J<0$, leading to an effectively reduced total moment especially
close to the ``quantum critical'' point. Note that this effect is enhanced when
one includes interactions in the band electron system (see e.g.\ \cite{Peters:2007}).

Away from half filling there appear, at strong enough doping, ferromagnetic phases
in addition \cite{santos:2002,Peters:2007}. 
For $J<J_c$, no further magnetic phase was observed in the vicinity
of half filling. For $J>J_c$, one can however again stabilize antiferromagnetic
phases. For example, for a filling $n_c=0.9$ of the conduction band, 
such an ordered state is found for $J\lesssim-W/4$. To obtain a reasonable
convergence and a stable solution, the DMFT calulations must be done
with Broyden mixing \cite{Zitko:2009Broyden}. Quite interestingly, we are not
able to find stable and reasonably converged solutions for
$-0.125\lesssim J/W\lesssim-0.1$. The  results for
\begin{figure}[htb]
  \begin{center}
\begin{minipage}[t]{0.6\textwidth}\mbox{}\\
  \includegraphics[width=0.99\textwidth]{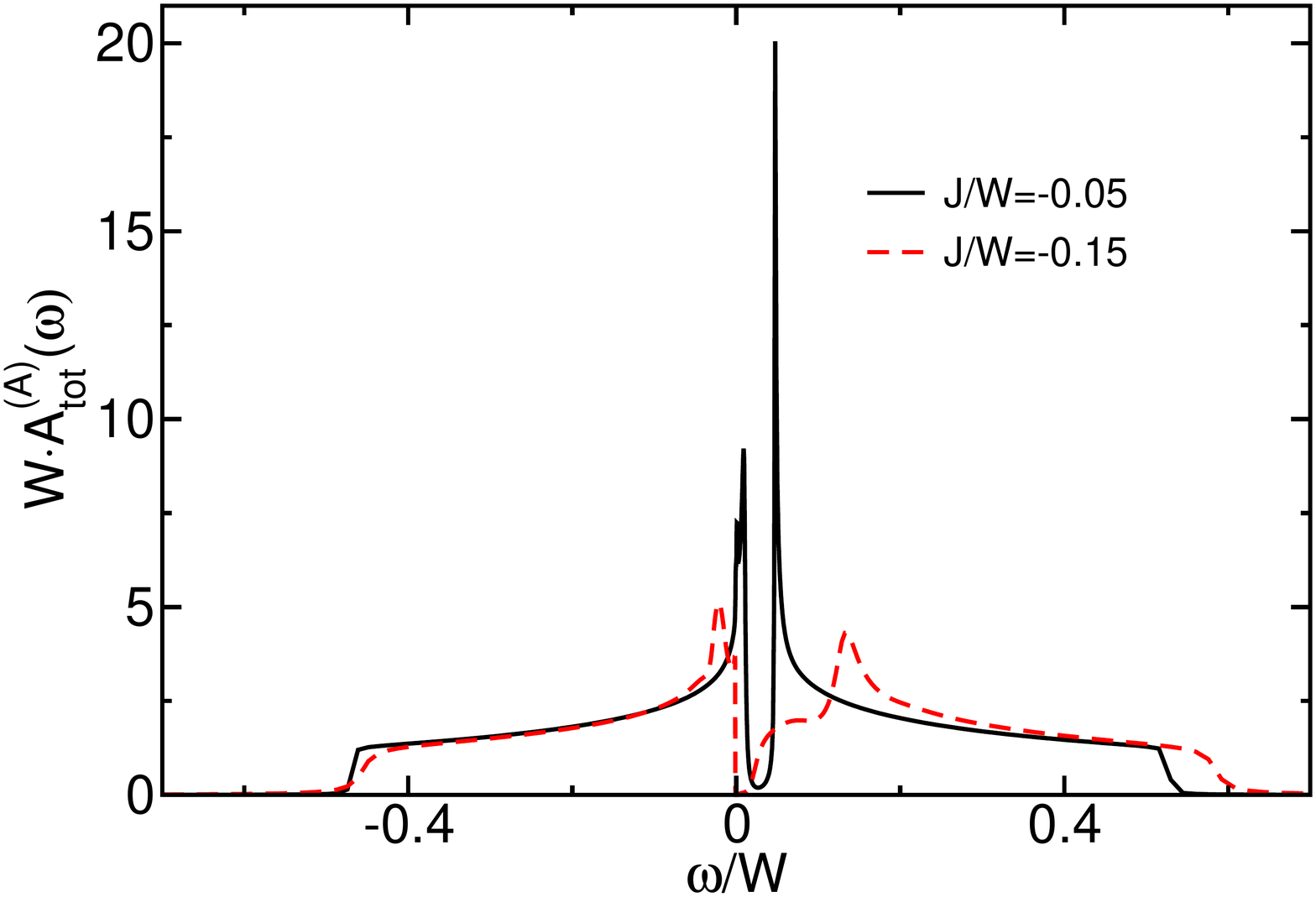}
\end{minipage}\hfill\begin{minipage}[t]{0.35\textwidth}\mbox{}\\
\begin{tabular}{@{}l@{\ }l@{\ }l@{\ }l}
\br
$J/W$ & $m_s^I$ & $m_s^c$ & $m_s^t$\\
\mr
$-0.05$ & $0.485$ & $0.057$ & $0.425$\\
$-0.15$ & $0.2$ & $0.08$ & $0.12$\\
\br
\end{tabular}
\end{minipage}
\end{center}
  \caption{Total DOS for the $A$ sublattice at $T=0$ for $J/W=-0.05$ (full line)
    and $J/W=-0.15$ (dashed line). The table to the left shows the staggered
    magnetizations.}
  \label{fig:DOS_AFM}
\end{figure}
$J/W=-0.05$ and $J/W=-0.15$ are shown in 
Fig.\ \ref{fig:DOS_AFM} together with a table of the values for the staggered 
magetization of the local
spin, $m_s^I$, the conduction band, $m_s^c$, and their sum, $m_s^t$.
%, are collected
%in the table.
%Tab.\ \ref{tab:staggered_magnetisation}.
%\begin{table}[htb]
%\caption{Staggered magnetization for $n_c=0.9$.}
%\label{tab:staggered_magnetisation}
%\begin{indented}
%\item[]
%\begin{tabular}{@{}l@{\ }l@{\ }l@{\ }l}
%\br
%$J/W$ & $m_s^I$ & $m_s^c$ & $m_s^t$\\
%\mr
%$-0.05$ & $0.485$ & $0.057$ & $0.425$\\
%-0.15$ & $0.2$ & $0.08$ & $0.12$\\
%\br
%\end{tabular}
%\end{indented}
%\end{table}
A rather interesting question in this connection is whether the ordered state
corresponds to a ``local moment'' regime, i.e.\ where the local spins are 
effectively decoupled from the band states and one has a small Fermi surface,
or if it is a ``heavy Fermion'' magnet with a large Fermi surface. The most
interesting case of course is when both appear as function of $J$ and there
might be a phase transition associated with the change in Fermi surface
topology. Obviously, inspection of the DOS in Fig.\ \ref{fig:DOS_AFM} alone is not sufficient  to identify a possible transition. 

\begin{figure}[htb]
\centerline{$J/W=-0.05$}
  \begin{minipage}{0.45\textwidth}
    \begin{center}
      \includegraphics[width=\textwidth]{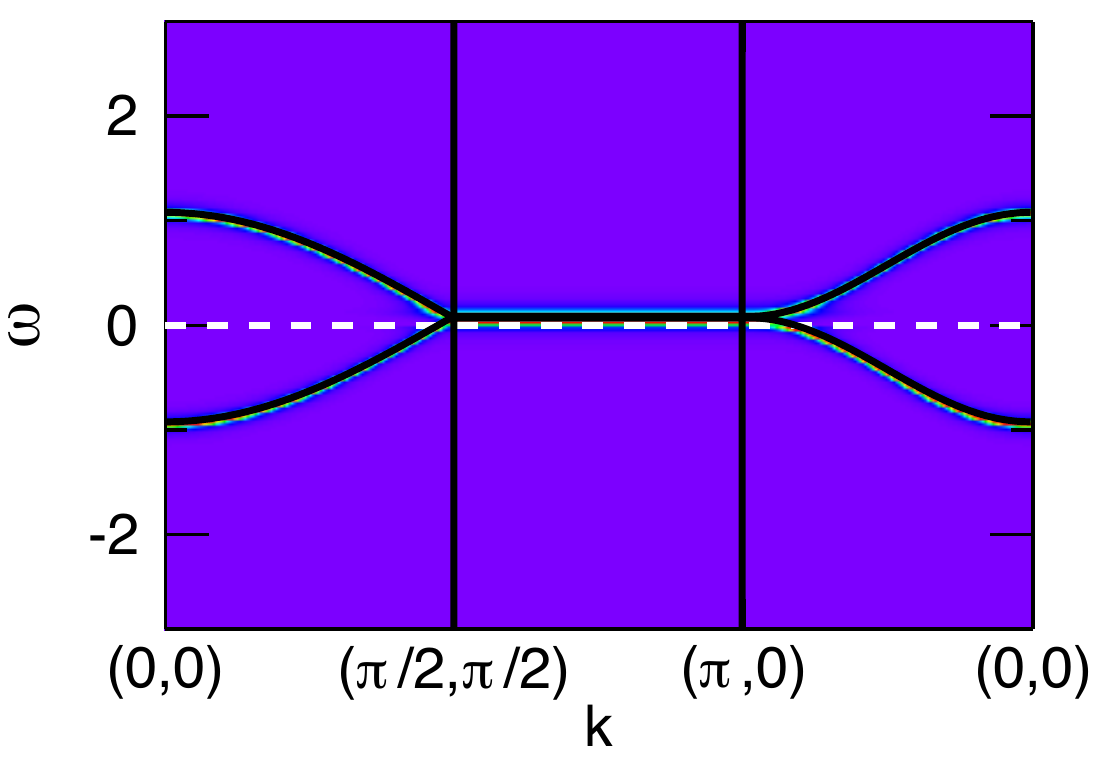}
    \end{center}
  \end{minipage}
  \hfill
  \begin{minipage}{0.45\textwidth}
    \begin{center}
      \includegraphics[width=\textwidth]{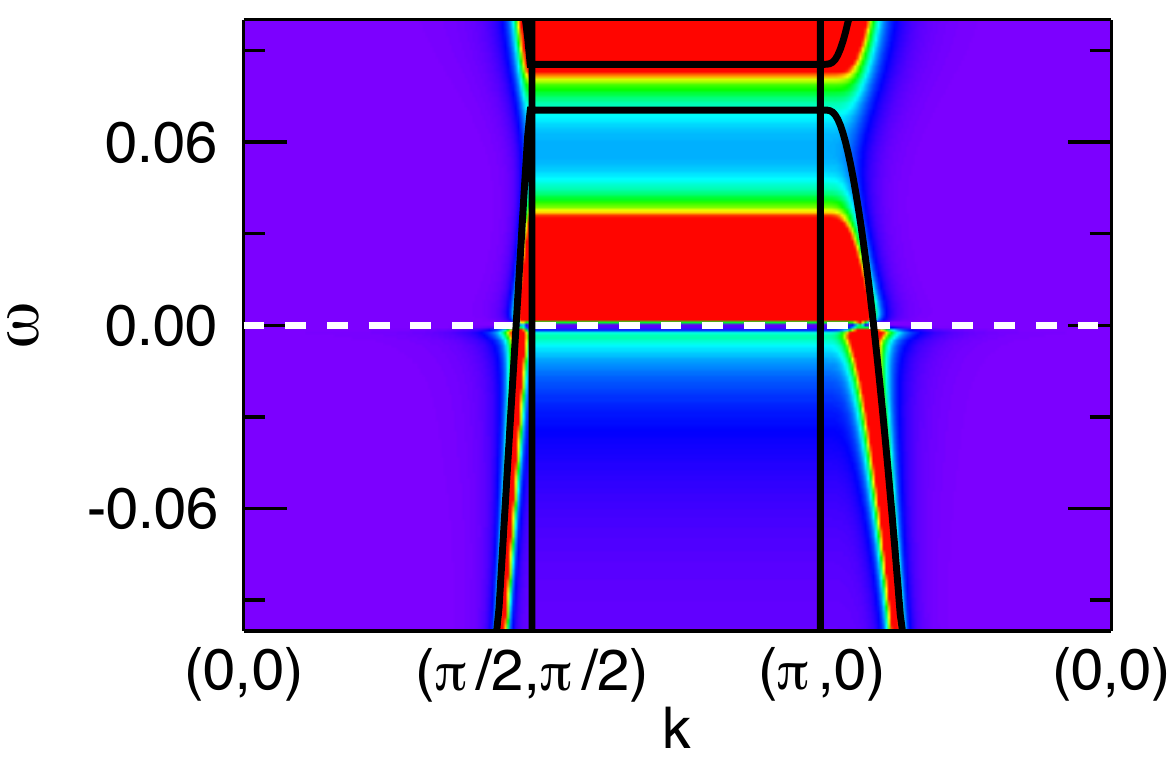}
    \end{center}
  \end{minipage}\\
\centerline{$J/W=-0.15$}
  \begin{minipage}{0.45\textwidth}
    \begin{center}
      \includegraphics[width=\textwidth]{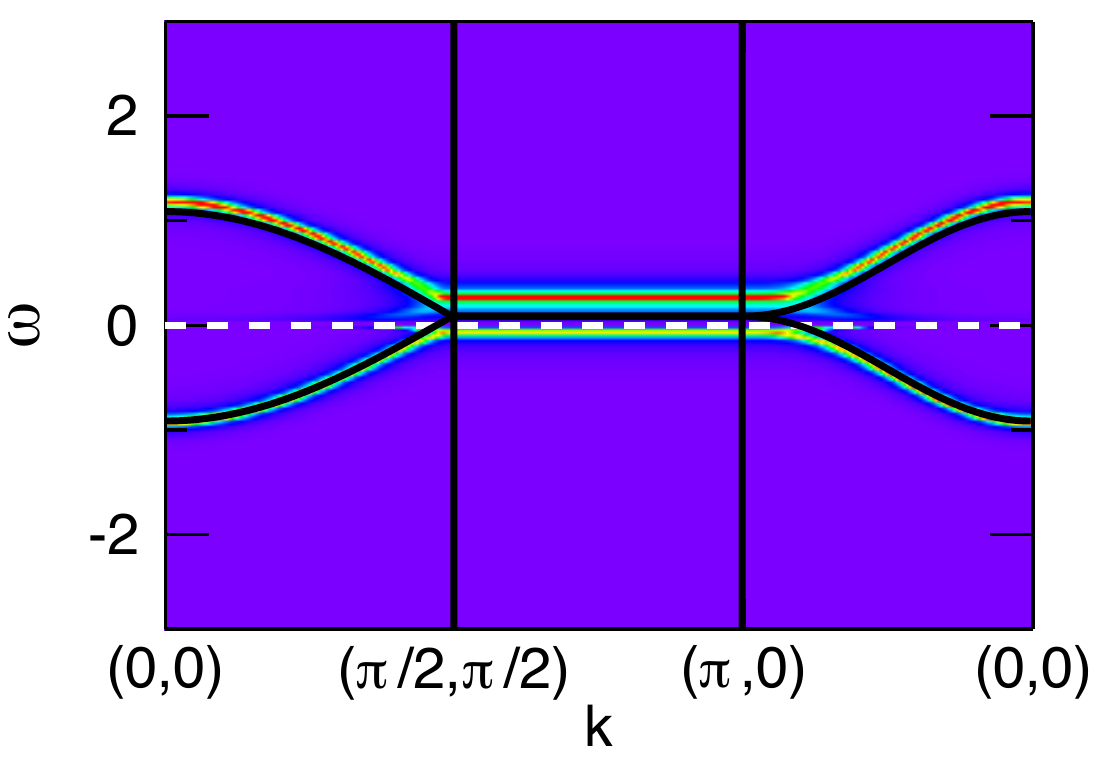}
    \end{center}
  \end{minipage}
  \hfill
  \begin{minipage}{0.45\textwidth}
    \begin{center}
      \includegraphics[width=\textwidth]{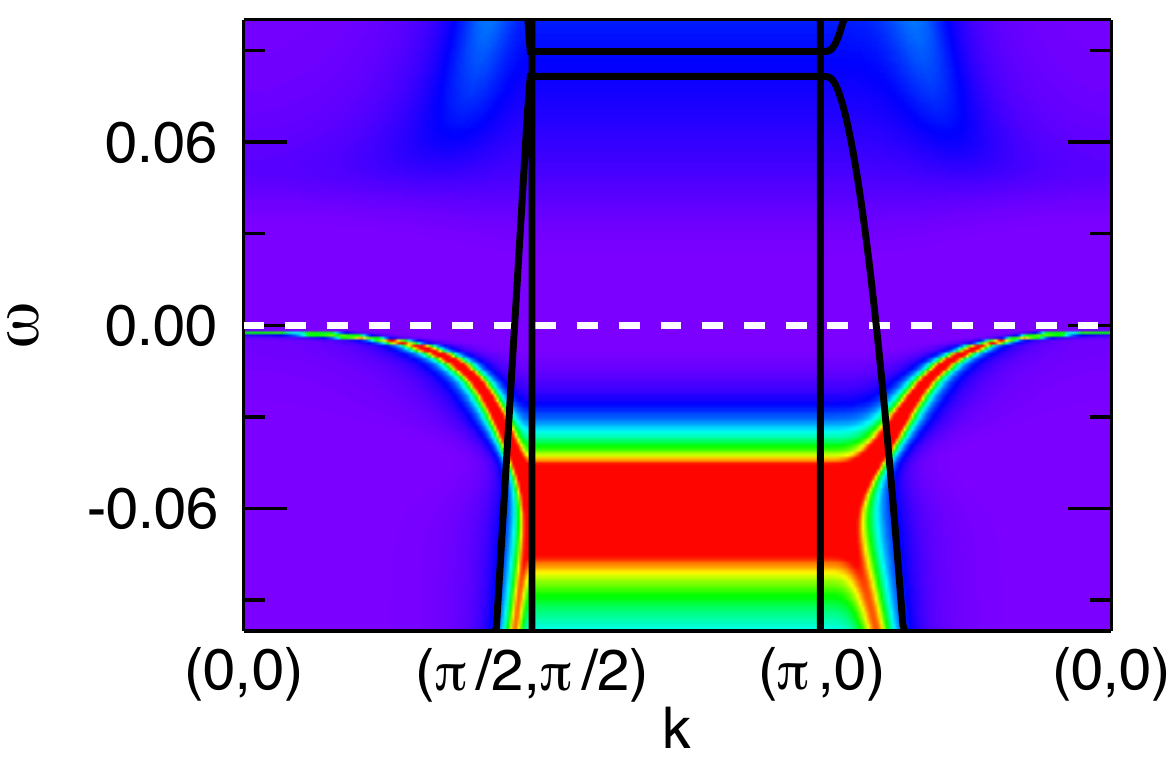}
    \end{center}
  \end{minipage}
  \caption{Spectral functions for $\vec{k}$ from the magnetic Brillouin zone
    for $n_c=0.9$. Left: Full energy range. Right: Magnified view around the 
    chemical potential.}
  \label{fig:Aofkw_AFM_small_J}
\end{figure}

Such information can however be obtained by inspecting the spectral function.
The results 
%for $J/W=-0.05$ and $n_c=0.9$ 
are shown in Fig.\ \ref{fig:Aofkw_AFM_small_J}, where the $\vec{k}$-vectors
are now restricted to the first magnetic Brillouin zone (see for example
\cite{pruschke:03}).
Included in the figure is as continuous black line the dispersion calculated 
with the Hartree term of the self-energy alone \cite{Negele}. This curve thus represents the 
proper approximation for the situation of
ordered local moments polarizing the Fermi sea.
The left part shows the overall structures and the right part a magnified view
of the region around the Fermi energy. 

Quite apparently, the concept of a 
small Fermi surface, i.e.\ ordered local spins polarizing the band states,
describes the situation very well for $J/W=-0.05$.
The situation changes completely for $J/W=-0.15$. 
%in Fig.\ \ref{fig:Aofkw_AFM_large_J}. 
First, the staggered magnetization has dropped to $m_s\approx0.12$ now, with $m_s^I\approx0.2$ and
$m_s^c\approx0.08$ (see Tab.\ \ref{fig:DOS_AFM}). Thus, while the polarization of the band states has
not changed significantly, the one for the local spins has dropped by
more than 50\%. Further increasing $|J|$ does not lead to any new features, the polarizations
of both local moments  and band electrons smoothly drop to zero for $J/W\approx-W/4$.
Second, the spectral function has become
% \begin{figure}[htb]
%   \begin{minipage}[t]{0.45\textwidth}
%     \begin{center}
%       $g=0.2W$\\
%       \includegraphics[width=\textwidth]{Aofkw_AFM_large_J}
%     \end{center}
%   \end{minipage}
%   \hfill
%   \begin{minipage}[t]{0.45\textwidth}
%     \begin{center}
%       $g=0.4W$\\
%       \includegraphics[width=\textwidth]{Aofkw_AFM_large_J_blowup}
%     \end{center}
%   \end{minipage}
%   \caption{Spectral function for $\vec{k}$ vectors from the magnetic Brillouin zone in the antiferromagnetically ordered state
%     for $J/W=-0.15$ and $n_c=0.9$ for the full energy range (left) and 
%     a magnified view around the chemical potential (right). 
%     High intensity is represented by red, low by 
%     violet. The black curve represents the bare dispersion split by the Hartree term.}
%   \label{fig:Aofkw_AFM_large_J}
% \end{figure}
much more HF like with a flat band around $\Gamma$ (c.f.\ Fig.\ \ref{fig:Aofkw_no_phonons}). Therefore, the 
scenario of a decoupled Fermi sea and local moments ordering does surely
not apply any more, in fact the whole system looks much more band like, with
a large Fermi surface.\footnote{Note that ``large'' in the magnetic Brillouin
zone means centered around $\Gamma$, as the $M$ point has been mapped back
to the zone center.} This scenario is also supported by the fact, that
for increasing coupling $|J|$ the system more strongly shows tendencies
to rather form spin density waves than N\'eel order. These observations strongly hint towards
a phase transition from a local moment like phase at small $|J|$  to a phase where magnetic ordering
appears in the heavy quasiparticles at larger $|J|$. However, a decisive answer what type of phase transition this is at present cannot be given. 

\section{Effect of phonons\label{sec:phonons}}
\begin{figure}[htb]
  \centering
  \includegraphics[width=0.8\textwidth]{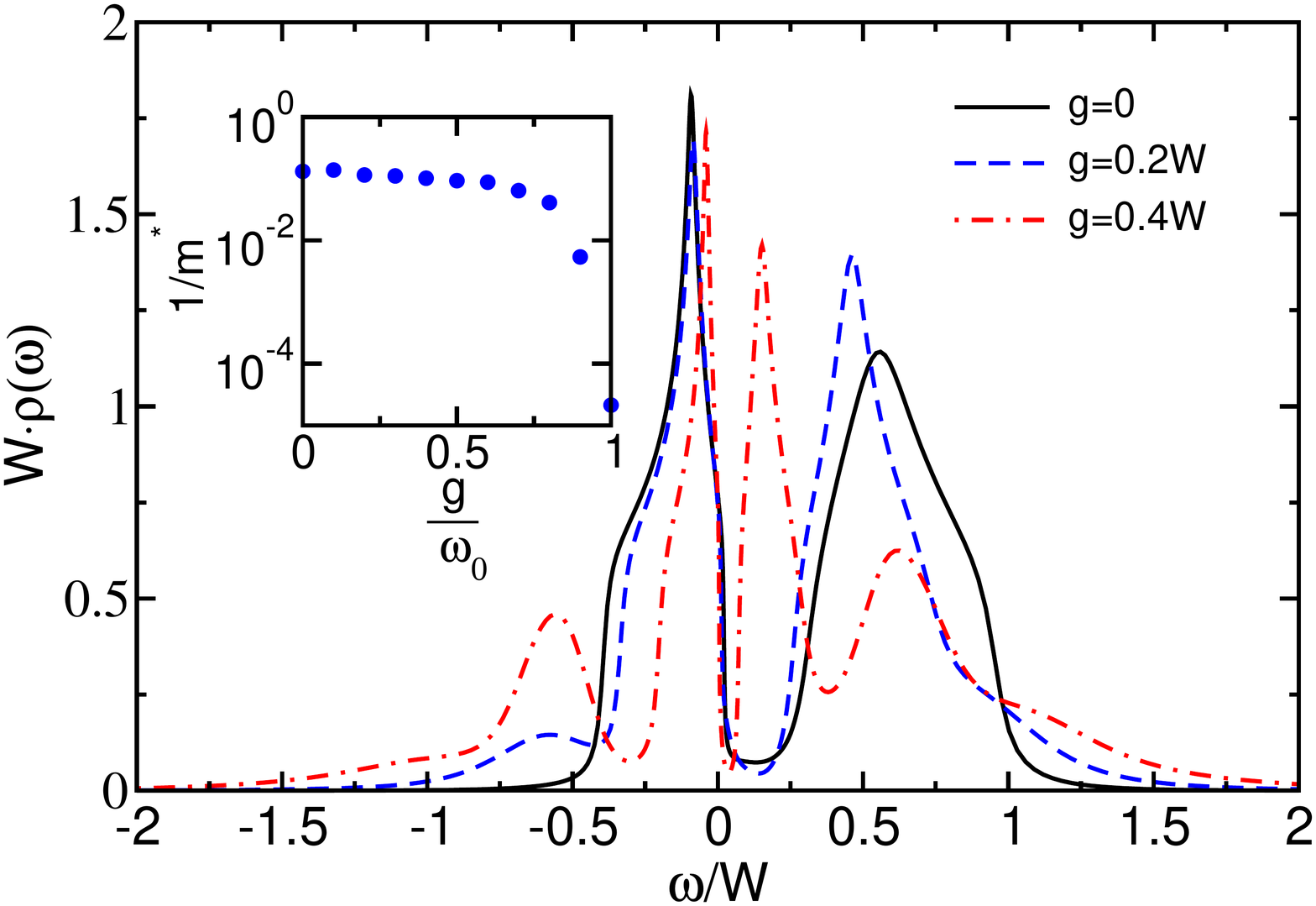}
  \caption{DOS at $T=0$ for the KLM \eqref{eq:KLM} with $J=-0.5W$ and
$g=0$, $g=0.2W$ and $g=0.4W$. The filling of the band is $n_c=0.8$.
The inset shows the dependence of the effective mass on $g$.}
  \label{fig:DOS_with_phonons}
\end{figure}
An extended account of the effect of Einstein phonons on HF physics
is given in \cite{assaad:2009}, where we concentrated on the periodic Anderson model.  Here, we show further results for the KLM, which supplement the findings reported there, in particular the behavior as the Kondo effect collapses due to polaron formation. We restrict the discussion here to 
$J<0$ and fix the phonon frequency to $\omega_0=0.5W$. Calculations were
done for $J=-0.5W$ at a filling $n_c=0.8$. The values for $\omega_0$ and $J$ 
are apparently rather large. However, smaller $J$ and $\omega_0$ do not change 
the qualitative
picture, but make it much harder to visualize the structures.

The results are summarized in Fig.\ \ref{fig:DOS_with_phonons} for $g=0$,
$g=0.2W$ and $g=0.4W$. 
The first thing to note is that the phonons lead 
to a reduction of the width of the pseudo gap close to the Fermi energy
and also a reduction of the overall bandwidth. In addition there occur
new structures at higher energy with increasing electron-phonon coupling,
which are related to the formation of polarons.
The reduction of the overall bandwidth is expected and can be interpreted as
an increase of the effective mass of the bare conduction states due to the
coupling to the phonons. The reduction
of the width of the pseudo gap, on the other hand, signals a likewise
reduction of the low-energy scale generated by the Kondo effect. Both effects
lead to an effective mass as function of $g$ as depicted in the inset to
Fig.\ \ref{fig:DOS_with_phonons}. Note that $m^\ast$ initially depends only
weakly on $g$. However it diverges very strongly as $g\to\omega_0$.

\begin{figure}[htb]
  \begin{minipage}[t]{0.45\textwidth}
    \begin{center}
      $g=0.2W$\\
      \includegraphics[width=\textwidth]{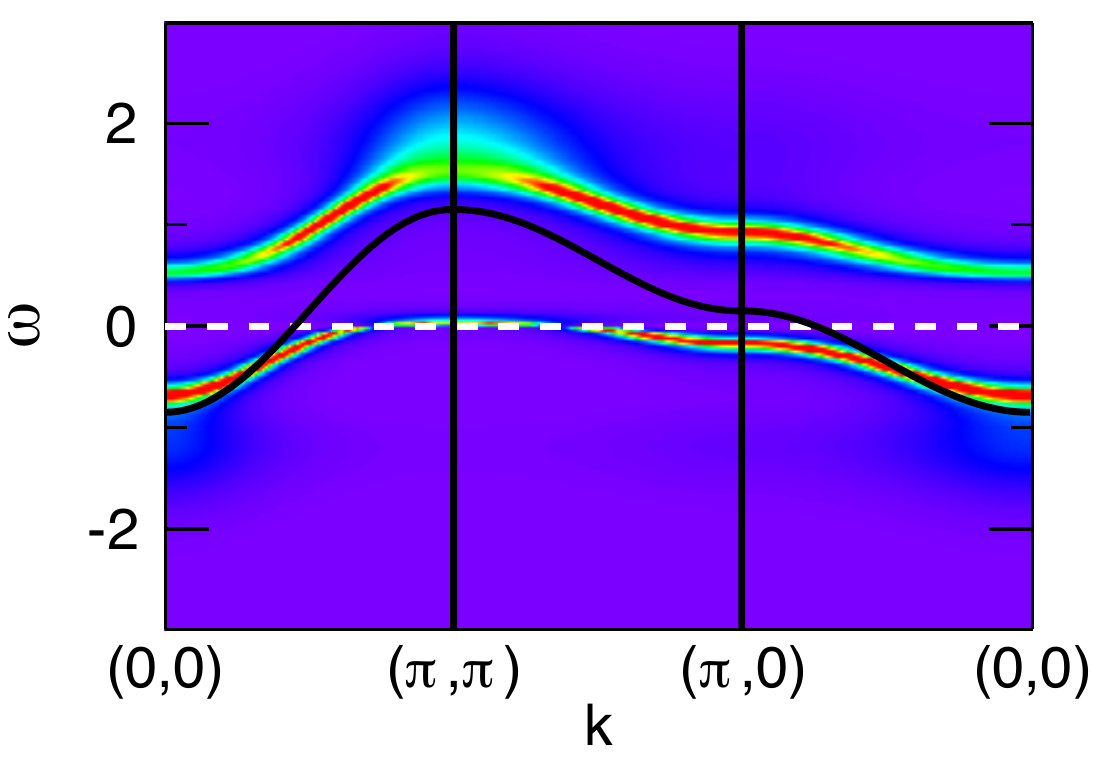}
    \end{center}
  \end{minipage}
  \hfill
  \begin{minipage}[t]{0.45\textwidth}
    \begin{center}
      $g=0.4W$\\
      \includegraphics[width=\textwidth]{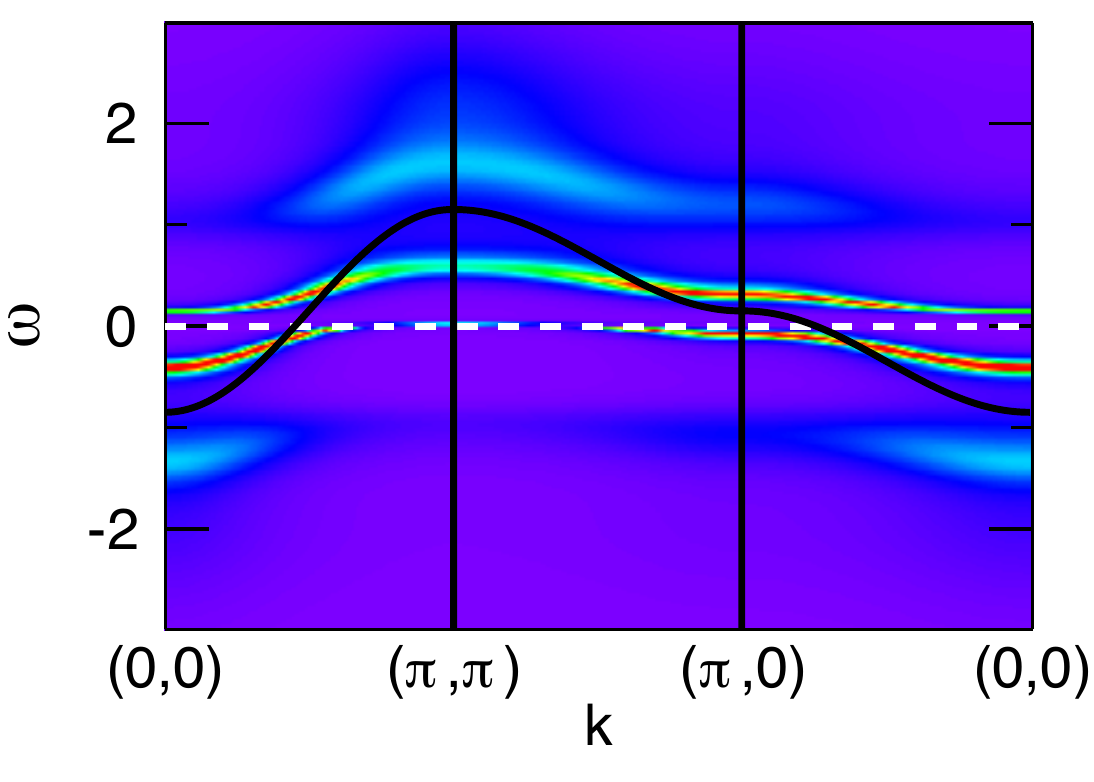}
    \end{center}
  \end{minipage}
  \caption{Spectral functions for $J=-0.5W$, other parameters as in Fig.\ \ref{fig:DOS_with_phonons}. High intensity is represented by red, low by violet.}
  \label{fig:Aofkw_with_phonons}
\end{figure}
These feature becomes more apparent by inspecting the spectral functions in Fig.\ \ref{fig:Aofkw_with_phonons}. Both effects, the overall reduction of the bare
bandwidth and the increased HF mass, are clearly visible. Moreover, with
increasing $g$ one finds roughly $\vec{k}$-independent structures representing
the polaronic modes. Further increasing $g$, we observe a rather sharp crossover 
\begin{figure}[htb]
  \begin{minipage}[t]{0.45\textwidth}
    \begin{center}
      \includegraphics[width=\textwidth]{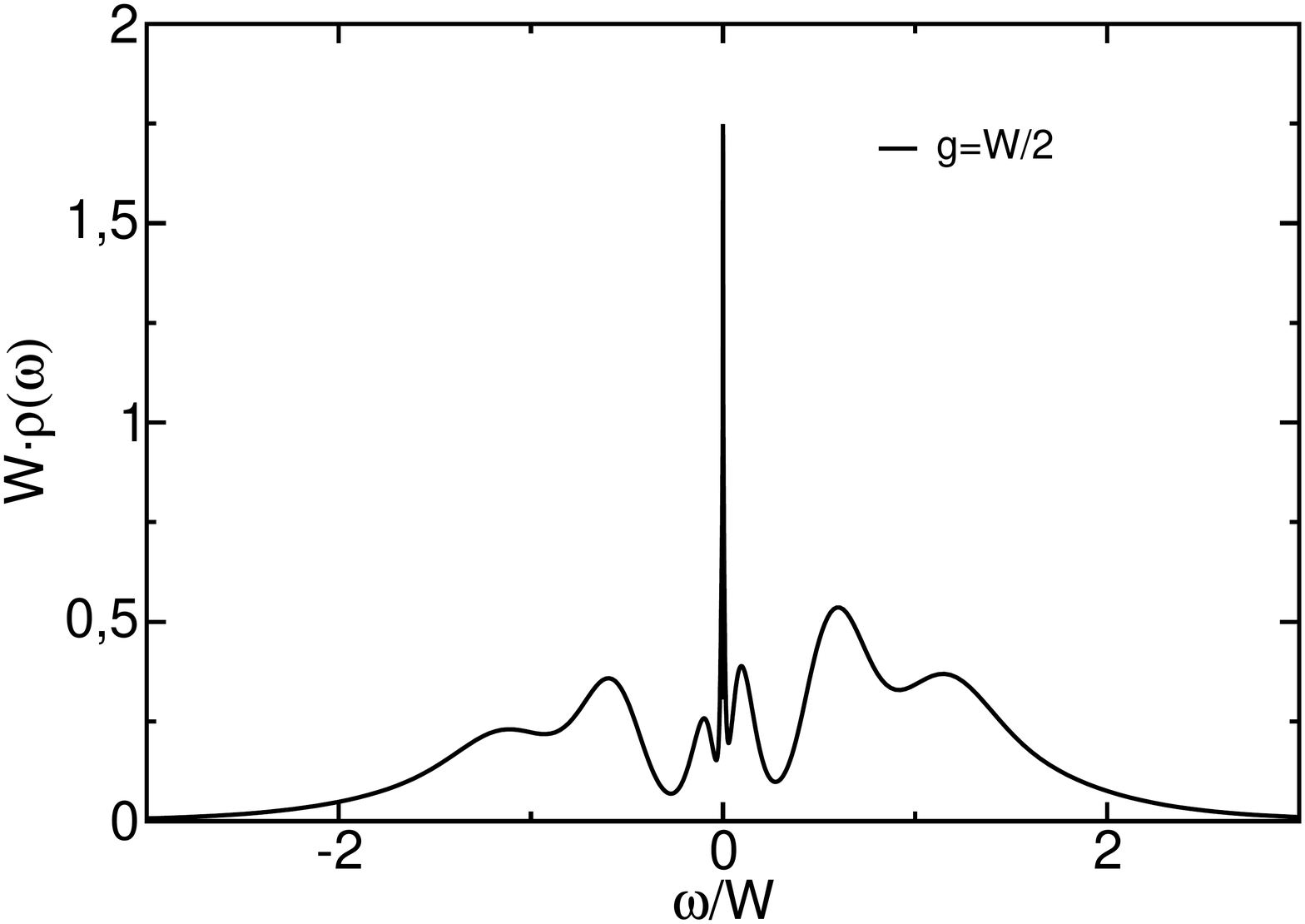}
    \end{center}
  \end{minipage}
  \hfill
  \begin{minipage}[t]{0.45\textwidth}
    \begin{center}
      \includegraphics[width=\textwidth]{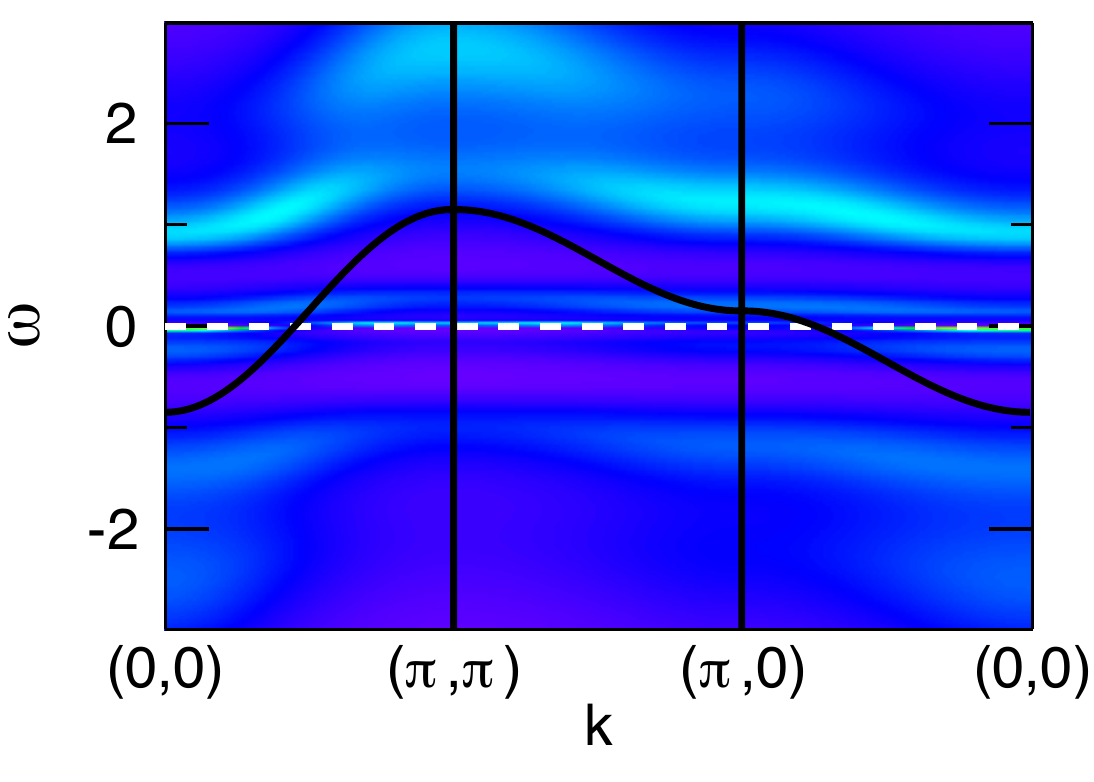}
    \end{center}
  \end{minipage}
  \caption{DOS (left panel)
respectively spectral function (right panel) for $g=0.5W=\omega_0$. Other parameters as in Fig.\ \ref{fig:DOS_with_phonons}.}
\label{fig:Phonons_large_g}
\end{figure}
around $g\approx \omega_0$ to a completely incoherent behavior. The results of a
calculation for $g=\omega_0$ are shown in Fig.\ \ref{fig:Phonons_large_g}.
Note that there is no Kondo feature left either in the DOS or in the spectral function,
and all structures are rather broad. We also would like to mention that for 
$g\gtrsim \omega_0$ it becomes increasingly hard to stabilize a given occupation
$n\ne0,1,2$ of the conduction band. This indicates that the system is close to a
Peierls instability, i.e.\ the formation of a charge density wave together with
a lattice distortion. 

\section{Summary}
We have presented a summary of properties of the Kondo lattice model within
dynamical mean-field theory at $T=0$ using the numerical renormalization group as
impurity solver. We have extended the Kondo lattice model by including
an Einstein mode coupled to the electrons via a Holstein term. 
The importance of such modes for HF materials can be deduced from 
strong effects such as the Kondo volume collapse observed in Ce.

Without the phonons, we find for antiferromagnetic
coupling the expected heavy-fermion behavior, with hybridized bands appearing in
the spectrum and a large Fermi surface. For ferromagnetic coupling on the other
hand, the bare band structure is only weakly modified due to incoherent scattering
from the local degrees of freedom.

Allowing for a magnetically ordered state, we are able to stabilize a
N\'eel order for $J>J_c$, where $J_c<0$ depends on the filling of the
conduction band. For the case $n_c=0.9$ discussed here, we find $J_c/W\approx-0.25$.
Quite interestingly, there appears to be a qualitative difference between
the ground-state for$|J|\to0$ and $|J|\to|J_c|$. In the former case, the
results can well be interpreted with a system consisting of local moments
which order antiferromagnetically via an RKKY-like exchange. In this case,
we have a small Fermi surface and the band states are weakly polarized by
the presence of the local moments. For the latter case we find a rather different
behavior. First, although the polarization of the local moments is considerably
smaller than for small $J$, the band polarization has actually increased pointing
towards a much stronger entanglement between local spins and conduction electrons. 
Second, the spectral functions again form flat bands around points of the magnetic
Brillouin zone which map to the heavy fermion bands without N\'eel order.
A more detailed investigation of the crossover respectively transitions could
not yet be accomplished due to convergence problems of the DMFT in the 
interesting region. 

Adding phonons, we find a general narrowing of the bare band, which also leads 
to a reduction of the Kondo scale. Eventually, when the coupling becomes
of the order of the phonon frequency, the electrons tend to localize and form
polarons with the phonons. At that point, the effective mass diverges and the
electronic spectrum becomes incoherent. As a side observation we note that
in this region one sees a tendency of the system to form a charge density wave.

% When we allow for superconductivity via a Nambu formulation, we find a 
% superconducting ground state at least for values $2g^2/\omega_0>T_0(g=0)$,
% where $T_0(g=0)$ represents the Fermi liquid scale of the system without phonons.
% This superconducting ground state is not BCS-like in the sense that the gap function,
% represented by the anomalous part of the self-energy, shows rather strong
% energy dependence. Moreover, the structure of the gap in the DOS is rather
% peculiar, showing asymmetries and additional features close to the gap edges.
% A similar behavior is also observed in tunneling experiments on high-$T_c$ compounds \cite{Lee:2009}.

The results for $T=0$ presented here strongly motivate further investigations,
in particular at finite $T$, searching for the critical temperature and
% looking into ratios like $\Delta(0)/T_c$. Here, $\Delta(0)$ can either be the
% limiting BCS value or it can be extracted from the gap in the DOS. Moreover, the present
% calculations were done for the antiadiabatic limit for numerical reasons. It
% is easy to extend them to real phonons, thereby including the additional 
% tendency towards polaron formation and charge ordering.
in particular the interplay between HF and local moment physics with respect
to magnetism, charge odering and superconductivity.

\ack
We want to thank helpful discussions with Andreas Honecker, Akihisa
Koga, Achim Rosch, Fakher Assaad and Dieter Vollhardt. 
RP wants to thank the Japan Society for the Promotion of Science
(JSPS) together with the Alexander von Humboldt-Foundation
 for a postdoctoral fellowship.
Computer support was provided by the Gesellschaft f\"ur
wissenschaftliche Datenverarbeitung in G\"ottingen and the
Norddeutsche Verbund f\"ur Hoch- und H\"ochstleistungsrechnen.

%\section*{References}
\bibliographystyle{jphysicsB}
\bibliography{references}
\end{document}